\documentclass[sigconf,screen,authorversion,nonacm]{acmart}

\usepackage{fancyhdr}
\AtBeginDocument{%
    \addtolength{\footskip}{2.0\baselineskip}
    \fancyfoot[L]{\textbf{\textit{© Philipp Danner 2025. This is the author's version of the work. It is posted here for your personal use. Not for redistribution. The definitive version was published in Proceedings of the 16th ACM International Conference on Future and Sustainable Energy Systems (E-ENERGY '25), \url{https://doi.org/10.1145/3679240.3734618}.}}}
}

\copyrightyear{2025}

\usepackage{pifont}
\usepackage{subcaption}
\usepackage[ruled,linesnumbered,vlined,commentsnumbered]{algorithm2e}

\SetCommentSty{mycommfont}
\newcommand{\cmark}{\ding{51}}
\newcommand{\xmark}{\ding{55}}
\DeclareMathOperator*{\argmax}{arg\,max}

\begin{document}

\title{Community Detection in Energy Networks based on Energy Self-Sufficiency and Dynamic Flexibility Activation}

\author{Philipp Danner}
\email{philipp.danner@uni-passau.de}
\orcid{0000-0002-3005-630X}
\affiliation{
  \institution{University of Passau}
  \city{Passau}
  \state{Bavaria}
  \country{Germany}
}

\author{Hermann de Meer}
\email{demeer@uni-passau.de}
\orcid{0000-0002-3466-8135}
\affiliation{
  \institution{University of Passau}
  \city{Passau}
  \state{Bavaria}
  \country{Germany}
}

\begin{abstract}
    The global energy transition towards distributed, smaller-scale resources, such as decentralized generation and flexible assets like storage and shiftable loads, demands novel control structures aligned with the emerging network architectures. These architectures consist of interconnected, self-contained clusters, commonly called microgrids or energy communities. These clusters aim to optimize collective self-sufficiency by prioritizing local energy use or operating independently during wide-area blackouts. This study addresses the challenge of defining optimal clusters, framed as a community detection problem. A novel metric, termed \textit{energy modularity}, is proposed to evaluate community partitions by quantifying energy self-sufficiency within clusters while incorporating the influence of flexible resources. Furthermore, a highly scalable community detection algorithm to maximize energy modularity based on the Louvain method is presented. Therefore, energy modularity is calculated using linear programming or a more efficient simulation-based approach. The algorithm is validated on an exemplary benchmark grid, demonstrating its effectiveness in identifying optimal energy clusters for modern decentralized energy systems.
\end{abstract}

\maketitle{}

\section{Motivation}
A major trend in the energy transition is the shift from large, centrally connected fossil-fueled power plants to a decentralized network of distributed energy resources, including photovoltaic, biomass, and wind power plants.
However, the inherent volatility of renewable energy generation necessitates flexibility resources to ensure grid stability and reliability. Those resources provide flexibility potentials, i.e., the controlled capability to deviate from a scheduled operating point~\cite{LECHL2023}. Examples are energy storage systems and sector-coupled, shiftable demands, such as heat pumps and electric vehicle chargers, which are increasingly embedded within the grid, lowering the need for grid expansion and thus reducing overall energy system cost~\cite{Brown2018}.
Activating flexibility potentials helps to balance supply and demand, ideally aligned as closely as possible in both spatial and temporal domains. Temporal alignment compensates for day-night and seasonal cycles in renewable generation to maximize self-sufficiency~\cite{danner2021flexibility}. On the other hand, spatial alignment minimizes energy flows over long distances, thus reducing transmission losses, and can help to alleviate grid congestion~\cite{Titz2024}.

New control structures have been proposed to match demand and supply at the most localized levels. These structures comprise interconnected, self-contained, and non-overlapping clusters (also called groups, modules, or communities) of resources with high internal energy density.
\textit{Microgrids}, typically implemented at medium or low-voltage levels, focus on electrical energy~\cite{lasseterMicroGrids2002,Smith2013}, while \textit{Energy Cells} extend this concept to integrate multiple domains, such as gas, heating, and mobility~\cite{Kiessling2011,uhlemeyer_cellular_2020}. \textit{Renewable Energy Communities}, as currently introduced in the European Union, focus on the social-technical integration of energy supply and demand within a regional context~\cite{Lowitzsch2020}.
By maximizing the self-sufficiency of each cluster, these structures can provide a first layer to pool and control flexibility resources. In an extreme case of wide-area blackouts, such structures could even enable temporary operation as energy islands~\cite{Wang2016a,Danner2024}.

Most municipalities in Europe, except for densely populated cities, could be self-sufficient concerning potential land use for energy supply and energy demand from inhabitants~\cite{trondle_home-made_2019}. Without limiting to municipal borders and by using the existing infrastructure, designing suitable control structures requires dividing the energy network into non-overlapping clusters of nodes. Each cluster should maximize its self-sufficiency while minimizing dependence on other clusters. Consequently, the problem of identifying such clusters can be formulated as a community detection problem, providing a systematic approach to partitioning the grid for optimal performance.

Related work covers the spatial domain using static structural information, a single snapshot of the system's operating state, e.g., after a fault, or a combination of both as input to community detection.
Handling the temporal domain is significantly more complex, and community detection problem formulations face several limitations: solving algorithms lack scalability, the number of communities needs to be predefined, and the impact of flexibility resource activation on self-sufficiency is frequently overlooked; cf. Table~\ref{tab:related_work_comparison}.

This work tackles the challenge of finding an optimal partition of energy networks with each community having maximum energy self-sufficiency using flexibility potentials and distributed renewable energy generation. Therefore, a community detection problem is formulated considering both temporal and spatial aspects. The community detection problem is linked with a self-sufficiency maximization by optimally using available flexibility potentials.
The core contributions are:
\begin{itemize}
    \item{} Definition of \textit{energy modularity} as a novel quality metric to evaluate an energy grid partition based on self-suf\-fi\-cien\-cy of communities. It integrates the benefits of flexibility activation within the communities.
    \item{} Community detection algorithm to maximize \textit{energy modularity} by combining the Louvain algorithm with community-level flexibility optimization, achieved through linear programming and a faster, simulation-based approach.
    \item{} Evaluation of the proposed approach applied to an exemplary power grid as typical energy network.
\end{itemize}

The remainder of the work is structured as follows: In Section~\ref{sec:related_work}, related work is discussed. Section \ref{sec:methodology} details the methodology including energy network modelling, \textit{energy modularity} and the Louvain-based community detection algorithm with integrated flexibility optimization. The proposed approach is evaluated with an exemplary power grid in Section \ref{sec:evaluation} and concluded in Section~\ref{sec:conclusion}.

\section{Related Work}\label{sec:related_work}
In literature, community detection (or graph partitioning if the number of clusters is given~\cite{fortunato_community_2010}) is often applied to power grids as typical energy networks. Different node relationship measures, clustering objectives and algorithms have been used, as listed in Table~\ref{tab:related_work_comparison} and described in the following.

\begin{table*}[bt]
    \caption{Related work on node relationships, objectives and algorithms for community detection in energy networks.}
    \label{tab:related_work_comparison}
    \footnotesize{}
    \centering{}
    \def\arraystretch{1.44}
    \begin{tabular}{p{0.73cm} p{2.62cm} p{3.70cm} p{2.19cm} c c c c c c}
        \toprule{}
        Ref.                                                                           & Node relationship                          & Objective                                                                 & Algorithm                               & \shortstack[c]{Network                                                                                                                              \\ structure} & \shortstack[c]{Operating \\ state} & \shortstack[c]{Time \\ domain} & \shortstack[c]{Energy \\ storage}  & \shortstack[c]{Cluster \\ count} & \shortstack[c]{Meshed \\ network}  \\
        \midrule{}
        \cite{chen_community_2015}                                                     & Topology                                   & Local similarity (variant of Katz index)                                  & Greedy agglomerative                    & \cmark{}                      & \xmark{} & \xmark{}                      & \xmark{}                      & \cmark{}                      & \cmark{} \\
        \cite{guerrero_evolutionary_2019}                                              & Topology                                   & Modularity                                                                & Genetic                                 & \cmark{}                      & \xmark{} & \xmark{}                      & \xmark{}                      & \cmark{}                      & \cmark{} \\
        \cite{Blumsack2009}                                                            & Electrical connectivity                    & Clustering tightness, cluster count, size and connectedness               & Greedy agglomerative, spectral, genetic & \cmark{}                      & \xmark{} & \xmark{}                      & \xmark{}                      & \cmark{}                      & \cmark{} \\
        \cite{Cotilla-Sanchez2013}                                                     & Elec.\ connectivity                        & Cohesiveness, between/intra-cluster connectedness, cluster count and size & k-means/genetic                         & \cmark{}                      & \xmark{} & \xmark{}                      & \xmark{}                      & \cmark{}                      & \cmark{} \\
        \cite{guerrero_community_2018}                                                 & Elec.\ connectivity                        & Modularity                                                                & Genetic                                 & \cmark{}                      & \xmark{} & \xmark{}                      & \xmark{}                      & \cmark{}                      & \cmark{} \\

        \cite{Zhao2019,eddin_novel_2023}                                               & Elec.\ coupling strength                   & Modularity                                                                & Newman fast, Louvain                    & (\cmark{})\textsuperscript{2} & \cmark{} & \xmark{}                      & \xmark{}                      & \cmark{}                      & \cmark{} \\

        \cite{wang_understanding_2022}                                                 & Elec.\ functional strength                 & Modularity                                                                & Newman fast                             & \cmark{}                      & \cmark{} & \xmark{}                      & \xmark{}                      & \cmark{}                      & \cmark{} \\
        \cite{li2005strategic,ding2012two,sanchez2014hierarchical,panteli2016boosting} & Power flow (absolute, average, apparent)   & Minimum ratio cuts                                                        & Spectral clustering                     & \cmark{}                      & \cmark{} & \xmark{}                      & \xmark{}                      & (\cmark{})\textsuperscript{1} & \cmark{} \\
        \cite{hasanvand2017spectral}                                                   & Apparent power flow                        & Minimum ratio cuts                                                        & Spectral clustering                     & \cmark{}                      & \cmark{} & (\cmark{})\textsuperscript{3} & \xmark{}                      & \cmark{}                      & \xmark{} \\
        \cite{han_intentional_2022}                                                    & Elec.\ edge betweenness                    & Maximum cut                                                               & Girvan-Newman                           & \cmark{}                      & \cmark{} & (\cmark{})\textsuperscript{4} & \xmark{}                      & \cmark{}                      & \cmark{} \\
        \cite{nassar_adaptive_2016}                                                    & Probabilistic power flow                   & Minimum cut (<5\% loading)                                                & Edge cut                                & \cmark{}                      & \cmark{} & \cmark{}                      & \xmark{}                      & \cmark{}                      & \xmark{} \\
        \cite{arefifar_supply-adequacy-based_2012}                                     & Probabilistic real and reactive power flow & Minimum cut                                                               & Tabu search                             & \cmark{}                      & \cmark{} & \cmark{}                      & (\xmark{})\textsuperscript{5} & \xmark{}                      & \xmark{} \\
        \cite{Barani2019}                                                              & Scenario-based power flow                  & Min.\ cut and supply-demand balance                                       & MILP solver                             & \cmark{}                      & \cmark{} & \cmark{}                      & \cmark{}                      & \xmark{}                      & \xmark{} \\
        \midrule{}
        \shortstack[c]{this                                                                                                                                                                                                                                                                                                                                                                                     \\ work}  & Energy exchange                            & Energy modularity                                                                                                                                                                                                                                         (self-sufficiency) & Louvain                                 & \cmark{}                      & \cmark{} & \cmark{}                      & \cmark{}                      & \cmark{}                      & \cmark{} \\
        \bottomrule{}
    \end{tabular}
    \textsuperscript{1}feasible but only applied in~\cite{sanchez2014hierarchical}, \textsuperscript{2}fully connected graph in \cite{Zhao2019}, but considers electrical admittance in edge weights, \textsuperscript{3}sensitivity analysis with different load flow situations, \textsuperscript{4}updates on state changes, \textsuperscript{5}cyclic state of charge constraints missing due to probabilistic power flow, no storage limit
\end{table*}

\subsection{Node Relationship}
Instead of the unweighted adjacency matrix~\cite{guerrero_evolutionary_2019,chen_community_2015}, which only represents the topology of a graph, one approach is to use the line admittance as electrical connectivity between buses in the power grid~\cite{Blumsack2009, Cotilla-Sanchez2013,guerrero_community_2018}. Both approaches focus only on structural information of the power grid and neglect its operating state.

The operating state of the power grid at a specific time can be observed from the power flow. In literature, the absolute power flow~\cite{li2005strategic} or the average active power flow~\cite{ding2012two,sanchez2014hierarchical} (average between incoming and outgoing power of a line to account for power losses) are used to define the relationship of two nodes.
Using apparent power flow as edge weight is proposed to consider both active and reactive power balance~\cite{panteli2016boosting,hasanvand2017spectral}.

A stronger focus on the functional community structure is provided by \citet{Zhao2019}. They define the relationship between buses, called electrical coupling strength, as the ratio of transmission capacity to the equivalent impedance.
Compared to previous approaches, their weights are built for a fully connected graph and are not limited to directly connected buses following the network structure. This, however, complicates partitioning into connected components, which is addressed but not guaranteed in~\cite{eddin_novel_2023}.
The functional structure is extended in~\cite{wang_understanding_2022} by adding a measure of supply-demand coupling based on the equivalent admittance between each load-generation pair and their maximum generation, transmission, and load capacity. Their community detection approach focuses on a ``generation follows demand'' scenario. Thus, dispatchable conventional generators have been assumed instead of variable renewable energy sources.

Regarding the time domain, \citet{hasanvand2017spectral} have at least conducted a sensitivity analysis to check the stability of the resulting partition with different operational states.
\citet{han_intentional_2022} define the importance of a power line by electrical edge betweenness, which is based on a combination of generator-load power, line impedance, and the importance of the line topology. They further update an initial partition after the operating state changes over time, yielding an optimal partition for the current grid state.

\citet{nassar_adaptive_2016} describe a dynamic approach to form microgrids with non-fixed boundaries on 16 typical scenarios within a year and model the probabilistic nature of generation and demand. The clusters are identified by minimum line cut with a power flow below 5\% of rated line loading and are updated depending on the expected line loading per scenario.
In contrast, stable grid partitioning over a longer period is applied in~\cite{arefifar_supply-adequacy-based_2012} and~\cite{Barani2019}.
In order to account for the varying nature of distributed generation, \citet{arefifar_supply-adequacy-based_2012} defined two probabilistic indices representing the real and reactive power of the lines in the partitioning process that minimizes the annual energy transfer between resulting microgrids.

\subsection{Objectives and Algorithms}
Finding a non-trivial partition, i.e., not all nodes in one community or each node in its own community, of non-overlapping communities in a graph is often an NP-hard or NP-complete problem for several objectives, such as minimizing ratio cuts, normalized cuts, or maximizing modularity~\cite{fortunato_community_2010,brandes_modularity_2008}. Therefore, heuristics are commonly used to extract the number of communities and their borders.

In the context of community detection, meta-heuristics, like tabu search~\cite{arefifar_supply-adequacy-based_2012} or genetic algorithms~\cite{Blumsack2009,guerrero_evolutionary_2019,Cotilla-Sanchez2013,guerrero_community_2018} have been used. The drawback of meta-heuristics is that their convergence behavior yields difficulties, especially in the case of an unknown number of communities. However, genetic algorithms have been found to be a good option to improve an initial partition from, e.g., spectral graph clustering~\cite{Blumsack2009}.

With a given (weighted) adjacency matrix, spectral graph clustering is a suitable and often applied method to find the minimum ratio cuts~\cite{chanSpectralKwayRatiocut1994}. It is based on dimensionality reduction in combination with a secondary clustering process. The secondary clustering ranges from defining \textit{k-centroids}~\cite{li2005strategic}, using \textit{k-means}~\cite{hasanvand2017spectral,Blumsack2009} or \textit{k-medoids}~\cite{ding2012two}, and applying \textit{hierarchical clustering}~\cite{sanchez2014hierarchical,panteli2016boosting}. A reasonable estimate for the number of communities \textit{k} can be inferred from the underlying spectral information~\cite{sanchez2014hierarchical}. Spectral graph clustering is an efficient and scalable approach for identifying suitable partitions.
Static relationships between nodes, such as admittance or pre-calculated power flow, can easily be represented as edge weights in a graph. However, incorporating the time domain and dynamic flexibility potentials activation to optimize a community's self-sufficiency presents a challenge. This is because the impact of flexibility activation on power flow between nodes depends dynamically on the demand and supply patterns specific to that community to be identified. As a result, spectral graph clustering methods are inadequate, as they cannot account for time dynamics and the effect of flexibility activation.

Other algorithms for community detection are based on modularity optimization, which is NP-complete~\cite{brandes_modularity_2008}. Modularity evaluates the quality of a graph partition by measuring how much more densely the nodes are connected within the proposed communities than they would be in a random distribution of edges.
Both divisive, such as the Girvan-Newman algorithm~\cite{han_intentional_2022}, and agglomerative algorithms, like the Newman fast algorithm~\cite{Zhao2019,wang_understanding_2022} or Louvain algorithm~\cite{eddin_novel_2023}, have been used for greedy modularity optimization in power grids. As with spectral graph clustering, the time domain and dynamic flexibility activation have not yet been integrated.

The approach in \cite{arefifar_supply-adequacy-based_2012} to find cuts on minimum power flows is based on probabilistic load flow, graph-related theories, and tabu search optimization. Storage systems are modelled as generators during the day, limited by power and energy constraints. Tracking the state of charge over time is missing and assumed to be feasible in a daily charge and discharge cycle.
\citet{Barani2019} partition radial power grids into microgrids using two distinct objectives: Minimizing the inter-microgrid power flow (both real and reactive power) and minimizing the demand-supply imbalance of constructed microgrids using a probabilistic model. The inter-microgrid power flow alone can result in infeasible islands, i.e., communities without generators. However, demand-supply imbalance provides better results.
Their mixed integer optimization approach is limited in scalability and applicable to radial grids only. The number of communities cannot be detected but needs to be predefined.

Integrating the time domain in complex networks can be modelled with dynamic graphs. They are characterized by changing structures over time, like the existence of nodes or edges or changes in their weights~\cite{harary_dynamic_1997}. Other than in social networks~\cite{alotaibi_review_2022}, nodes and edges in the power grid, which are buses and power lines, respectively, are unlikely to appear and disappear over time (except for transient changes in abnormal situations or permanent, long-term changes due to grid reinforcement once in several years~\cite{fletcher2007optimal}). However, edge weights are directly linked to demand and supply patterns and change over time. Furthermore, flexibility activation directly impacts the edges by changing their power flow.
A comprehensive survey on algorithms for community detection in dynamic graphs can be found in~\cite{rossetti_community_2019}. A big challenge of community detection in dynamic graphs is obtaining a stable community partition over time. In the application to power grids, communities are therefore either retrieved and updated iteratively on changes~\cite{han_intentional_2022,nassar_adaptive_2016} or by considering all temporal information at once~\cite{arefifar_supply-adequacy-based_2012,Barani2019} to get a stable community structure. This work focus on the latter because flexibility potential constraints cannot be respected when modelling each time slice separately.

Related work often uses structural information or a single snapshot of the system's operating state, e.g., after a fault. Handling the time domain is much more complex, and community detection problem formulations and their solving algorithms are limited in scalability, require the number of communities to be predefined, can only be applied to radial networks, and/or neglect the effect of flexibility activation on the power flow between nodes, see Table~\ref{tab:related_work_comparison}.
In this work, a community detection algorithm is designed to optimize a novel energy self-sufficiency metric inspired by modularity to assess the quality of a partition. This metric is called energy modularity. The time domain and flexibility resources, modeled as energy storage systems, are considered. By employing a Louvain-based algorithm, communities can be detected within a reasonable timeframe while considering flexibility activation.

\section{Community Detection}\label{sec:methodology}
This section outlines a novel approach to identifying energy self-sufficient communities in energy networks.

\subsection{Problem Formulation}\label{sec:problem_formulation}
In the context of this work\footnote{Upper-case letters are used for matrices; Non-cursive letters are parameters, \textit{cursive} letters are variables.}, an energy network is modelled as a directed, weighted graph \( \mathcal{G} = (\mathcal{V}, \mathcal{E}) \) with an edge weight function \( \omega{}: \mathcal{E} \to{} \mathbb{R}_{\geq{}0} \). The node set \( \mathcal{V} \) represents the energy network nodes, and the edge set \( \mathcal{E} \subseteq{} \{(u,v)\;|\;u,v \in \mathcal{V}\} \) defines the direct physical connections and thus the capability to carry \textit{energy flows} from node \( u \) to \( v \). In general, energy exchange is bidirectional, thus edges in both direction exist, i.e., \( (u,v)\in{\mathcal{E}}\) and \( (v,u) \in{\mathcal{E}} \).

\subsubsection{Edge Weights and Energy Flows}
\begin{figure}[tb]
    \centering{}
    \includegraphics{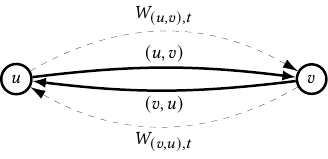}
    \caption{Energy graph (bold) with energy flows (dashed)}
    \label{fig:graph}
    \Description{Two nodes \( u \) and \( u \) are shown with bidirectional, directed edges. On each edge, an energy flow for time \( t \) is given.}
\end{figure}

Each edge carries energy flows for the time set \( \mathcal{T} = \{1, \dots{},\mathrm{h}\} \) of sequentially ordered time slices up to the horizon \( \mathrm{h} \), see Figure~\ref{fig:graph}.
An energy flow on an edge \( e \in{\mathcal{E}} \) at time \( t \) is given by  \( W_{e,t} > 0 \) with \( W \in{\mathbb{R}_{\geq{}0}^{\mathcal{E}\times{}\mathcal{T}}} \) being a discrete map \( \mathcal{E} \times \mathcal{T} \to{} \mathbb{R}_{\geq{}0} \). Energy flows are limited by \( \overline{\mathrm{w}} \in{\mathbb{R}_{\geq{}0}^{\mathcal{E}}} \):
\begin{align}\label{eq:energy_flow_limit}
    W_{e,t} \; \leq{} \; \overline{\mathrm{w}}_{e} \quad{} \forall{e\in{\mathcal{E}},t\in{\mathcal{T}}}\,.
\end{align}
The edge weight function \( \omega{} \) quantifies the total energy flows on a given edge \( e \in{\mathcal{E}} \) over time and is defined by:
\begin{align*}
    \omega{}\bigl(e\bigr) \; = \; \sum_{t\in{\mathcal{T}}} W_{e,t}\,.
\end{align*}

\subsubsection{Node Characteristics}
Nodes have time-varying energy \textit{demand} and \textit{supply}, which are given in the parameters \( \mathrm{D}, \mathrm{S} \in{\mathbb{R}_{\geq{}0}^{\mathcal{V} \times \mathcal{T}}} \), respectively.
Furthermore, each node has a flexibility potential, modelled as an energy storage system. With flexibility potential activation, demand or supply can be shifted within \( \mathcal{T} \), i.e., by storing energy for later use. The yet unknown \textit{positive flexibility} \( F^{+} \in{\mathbb{R}_{\geq{}0}^{\mathcal{V} \times \mathcal{T}}} \), i.e., dispatch energy to the network, and \textit{negative flexibility} \( F^{-} \in{\mathbb{R}_{\geq{}0}^{\mathcal{V} \times \mathcal{T}}} \), i.e., consume and store energy from the network, have an upper limit in \( \mathrm{\overline{f}} \in{\mathbb{R}^\mathcal{V}} \):
\begin{align}\label{eq:f_max}
    F^{+}_{v,t} \; \leq{} \; \mathrm{\overline{f}}_{v}, \quad{} F^{-}_{v,t} \; \leq{} \; \mathrm{\overline{f}}_{v}
    \quad{} \forall{v\in{\mathcal{V}},t\in{\mathcal{T}}}\,.
\end{align}
Flexibility activation is constrained by its \textit{state of charge} \( SOC \in{\mathbb{R}_{\geq{}0}^{\mathcal{V} \times \mathcal{T}}} \) in its maximum capacity \(  \mathrm{\overline{soc}} \in{\mathbb{R}_{\geq{}0}^{\mathcal{V}}} \):
\begin{align}\label{eq:SOC_minmax}
    {SOC}_{v,t} \; \leq{} \; \mathrm{\overline{soc}}_{v}
    \quad{} \forall{v,t} \,.
\end{align}
Further, time-dependency is modelled by:
\begin{align}\label{eq:SOC_time}
    {SOC}_{v,t} \; = \;{SOC}_{v,t-1}\cdot{}\upeta^\mathrm{p}_{v} \;+\; F^{-}_{v,t}\cdot{}\upeta^\mathrm{u}_{v} \;-\; F^{+}_{v,t}\,/\,\upeta^\mathrm{u}_{v}
    \quad{} \forall{v,t} \,,
\end{align}
where \( \upeta^\mathrm{u} \in{{[0,1]}^{\mathcal{V}}} \) is the usage efficiency for positive and negative flexibility and \( \upeta^\mathrm{p} \in{{[0,1]}^{\mathcal{V}}} \) the energy-preserving efficiency for self-discharge effects. Flexibility resources should not dispatch more energy than is stored over time, given by the cyclic constraints:
\begin{align}\label{eq:SOC_cyclic}
    {SOC}_{v,0} \; = \; {SOC}_{v,\mathrm{h}}
    \quad{} \forall{v} \,.
\end{align}
This also models the rebound effect that occurs when flexibility is activated.
If no flexibility is available at node \( v \), \( \overline{\mathrm{f}}_{v} = \overline{\mathrm{soc}}_{v} = 0 \).

Flexibility potentials should support the community to which they belong. Therefore, flexibility activation must be determined individually for each community (see Section~\ref{sec:partition_quality} for more details).

\subsubsection{Nodal Energy Balance}
Nodal balance must be given according to the principle of conservation of charge (Kirchhoff's Current Law). Energy injected at each node by supply and positive flexibility, minus energy withdrawn from demand and negative flexibility, is the same as the sum of in and negative outgoing energy flows:
\begin{align}\label{eq:nodal_balance}
    \mathrm{D}_{v,t} + F^{-}_{v,t} - \mathrm{S}_{v,t} - F^{+}_{v,t} \; = \; \sum_{e\in{\mathcal{E}}} \mathrm{K}_{e,v} \cdot{} W_{e,t}
    \quad{} \forall{v,t}\,,
\end{align}
where \( \mathrm{K}_{e,v} \) is the incidence matrix regarding energy flows \( W_{e,t} \) which is \( \upeta^\mathrm{f}_{e} \) if edge \( e \) starts at node \( v \), \( -1 \) if edge \( e \) ends at node \( v \), and zero otherwise; \( \upeta^\mathrm{f} \in{\mathbb{R}_{\geq{}0}^{\mathcal{E}}} \) is the edge efficiency.

This formulation is known as the transport or network flow model with linear loss approximation, which is often used in network planning studies~\cite{neumannAssessmentsLinearPower2022}.
The actual distribution of energy flows in meshed networks, such as according to the impedance of lines in the power grid, is not considered, thus allowing for cyclic energy flows. Because the amount of energy flows is not relevant for community detection in energy networks but energy self-sufficiency, as discussed in Section~\ref{sec:partition_quality}, hypothetical cyclic flows are acceptable within this context.

\subsubsection{Communities and Clustering}
A community \( \mathcal{C} \subseteq \mathcal{V} \) is a unique subset of nodes. A set of \( k \) communities \( \mathcal{P} = \{ \mathcal{C}_1, \dots{}, \mathcal{C}_k \} \) is called a partition of the graph if the communities are non-over\-lap\-ping \( \mathcal{C}_i \bigcap{} \mathcal{C}_j = \emptyset \quad \forall i,j\in{\{1,\dots{},k\}}, i \neq j \) and the union of all \( k \) communities is the set of nodes \( \mathcal{V} = \bigcup{}_{i=1}^{k} \mathcal{C}_i \).
The goal is to find a partition \( \mathcal{P} \) of the graph \( \mathcal{G} \) so that a quality function \( Q \) to assess the goodness of the partition is maximized:
\begin{align}\label{eq:opt_partition}
    \mathcal{P} \;=\; \argmax_{\mathcal{P}} Q (\mathcal{P})\,.
\end{align}

\subsection{Partition Quality}\label{sec:partition_quality}
A quality function can be used to assess the goodness of a partition. When employing such a quality function as an objective in an optimization problem, it should not favor trivial partitions, such as placing all nodes in a single community or assigning each node to its own separate community. Instead, it should promote strong connections within communities and weak connections between them, as these characteristics intuitively indicate a good partition~\cite{traag_narrow_2011}.

\subsubsection{Standard Modularity}
A widely used quality function, even though not distribution limit free~\cite{traag_narrow_2011}, is \textit{modularity}, first introduced by~\citet{Newman2004}. It compares actual edge density within a community to the expected density if nodes are connected independently of any structure. Standard modularity can be expressed as:
\begin{align*}
    Q (\mathcal{P}) \; = \; \sum_{\mathcal{C}\in\mathcal{P}} \Bigl(e^{*} (\mathcal{C}) - {a^{*} (\mathcal{C})}^2\Bigr) \,,
\end{align*}
where \( e^{*} (\mathcal{C}) \) denotes the fraction of edges in the graph that lie entirely within community \( \mathcal{C} \), and \( a^{*} (\mathcal{C}) \) represents the fraction of edges that are connected to at least one node in \( \mathcal{C} \), including those linking to other communities. A weighted extension of this formulation incorporates edge weights instead of just the existence of an edge, therefore capturing more than only topological information.

There are two issues when applying weighted standard modularity for energy networks with energy flows as edge weights and flexibility potentials:
\textit{First}, energy flows are affected by the flexibility activation through \( F^{+} \) and \( F^{-} \) in the nodal balance constraint in Eq.~\eqref{eq:nodal_balance}, resulting in potentially network-wide edge weight changes. This interdependency eliminates a key advantage of (weighted) standard modularity: the ability to compute modularity contribution independently for each community, which is crucial for efficient heuristics, such as with the Louvain algorithm.

\textit{Second}, maximizing weighted standard modularity strives to increase energy flows within communities. However, the intuitive objective in energy networks is to maximize energy self-sufficiency within each community, a related but essentially different goal.
Maximizing edge weights can lead to unintended outcomes: Consider a single time slice \( t \) of an energy graph where a community has an undersupply of \( x \) units. Weighted standard modularity may favor adding a node \( v \) with \( F^{+}_{v,t} < x \) that connects through multiple edges, artificially inflating energy flows and edge weights. In contrast, a node \( w \) with \( F^{+}_{w,t} = x \) that connects directly to the energy-demanding node would be overlooked even though it would provide a more effective solution for addressing the undersupply.

\subsubsection{Energy Modularity}
Inspired by the structure of standard modularity, \textit{energy modularity} quantifies the degree of energy self-sufficiency within each community, while simultaneously promoting a higher number of communities to avoid the trivial case where all nodes are grouped into a single community.

The self-sufficiency of a community \( \mathcal{C} \) is determined by the portion of each node's demand that is supplied from within the same community. Specifically, \( \widetilde{D}_{w,t} \) denotes the share of demand at node \( w \in \mathcal{C} \) and time \( t \) that is covered by community-internal resources, while \( \widetilde{S}_{w,t} \) represents the corresponding internal supply. These quantities are constrained by the maximum nodal demand \( \mathrm{D}_{w,t} \) and supply \( \mathrm{S}_{w,t} \) as follows:
\begin{align}\label{eq:d_tilde}
    \widetilde{D}_{w,t} \; \leq{} \; \mathrm{D}_{w,t} \quad{}\forall{w\in{\mathcal{C}}, t\in{\mathcal{T}}}\,,
\end{align}
\begin{align}\label{eq:s_tilde}
    \widetilde{S}_{w,t} \; \leq{} \; \mathrm{S}_{w,t} \quad{}\forall{w\in{\mathcal{C}}, t\in{\mathcal{T}}}\,.
\end{align}
Further, the nodes in community \( \mathcal{C} \) and internal edges \( \mathcal{E}' \) are virtually cut out from the graph. The resulting community-internal nodal balance constraints, adapted from Eq.~\eqref{eq:nodal_balance}, are given by
\begin{align}\label{eq:d_opt_nodal_balance}
    \begin{split}
        \widetilde{D}_{w,t} + F^{-}_{w,t} - \widetilde{S}_{w,t} - F^{+}_{w,t} = \sum_{e\in{\mathcal{E}'}} \mathrm{K}_{e,w} \cdot{} W_{e,t} \\
        \forall{w\in{\mathcal{C}}, t\in{\mathcal{T}}}, \mathcal{E}' = \{(u,v)\;|\;(u,v) \in{\mathcal{E}}, u \in{\mathcal{C}}, v \in{\mathcal{C}}\}\,.
    \end{split}
\end{align}
The total demand \( d(\mathcal{C}) \), which is supplied from within community \( \mathcal{C} \), depends on the activation of flexibility potentials \( {F}^{+} \) and \( {F}^{-} \), and can be calculated by:
\begin{align}\label{eq:d_opt}
    \begin{split}
        d (\mathcal{C}) \; = \; &\max_{F^{+},F^{-}} \sum_{t\in\mathcal{T}} \sum_{w\in\mathcal{C}} \widetilde{D}_{w,t} \,, \\
        \text{s.t. }  \eqref{eq:energy_flow_limit}\text{,}\eqref{eq:d_tilde}\text{--}\eqref{eq:d_opt_nodal_balance} & \quad{} \text{ (community nodal balance)}\,, \\
        \eqref{eq:f_max}\text{--}\eqref{eq:SOC_cyclic} & \quad{} \text{ (flexibility)} \,. \\
    \end{split}
\end{align}

Similar to \( e^{*} (\mathcal{C}) \) in \textit{standard modularity}, the term \( e (\mathcal{C}) \) represents the fraction of community's demand that is met by internal supply, i.e., \( d(\mathcal{C}) \), normalized to the total demand of the whole network:
\begin{align}\label{eq:e_c}
    e (\mathcal{C}) \; = \; \frac{d (\mathcal{C})}{\sum_{t\in\mathcal{T}} \sum_{v\in\mathcal{V}} \mathrm{D}_{v,t}} \,.
\end{align}
Correspondingly, \( a(\mathcal{C}) \) denotes the fraction of total system demand that is attributable to community \( \mathcal{C} \), including demand supplied from other communities. This formulation adheres to the nodal balance constraint in Eq.~\eqref{eq:nodal_balance}, which ensures that all demands in the system are supplied by some node. The expression for \( a(\mathcal{C}) \) is given by:
\begin{align}\label{eq:a_c}
    a (\mathcal{C}) \; = \; \frac{ \sum_{t\in\mathcal{T}} \sum_{w\in\mathcal{C}} \mathrm{D}_{w,t} }{ \sum_{t\in\mathcal{T}} \sum_{v\in\mathcal{V}} \mathrm{D}_{v,t}} \,.
\end{align}
Energy modularity is defined as the sum of the contributions from each community, as given by:
\begin{align}\label{eq:q}
    Q (\mathcal{P}) \; = \; \sum_{\mathcal{C}\in{\mathcal{P}}} Q^{c} (\mathcal{C})\,,
\end{align}
\begin{align}\label{eq:q_c}
    Q^{c} (\mathcal{C}) \; = \; e (\mathcal{C}) - \gamma{} \cdot{} {a (\mathcal{C})}^2 \,,
\end{align}
where \( \gamma{} \) is a resolution parameter to favor larger communities with \( \gamma{} < 1 \) and smaller ones with \( \gamma{} > 1 \), following~\cite{reichardtPartitioningModularityGraphs2007a}.

Energy modularity inherits some useful properties from standard modularity. For \( \gamma = 1 \), the energy modularity of the trivial partition, where all nodes belong to a single community, is zero since \( e(\mathcal{V}) = a(\mathcal{V}) = 1 \). With an increasing number of entirely self-sufficient communities, energy modularity approaches 1. Thus, maximizing energy modularity yields an optimal grouping of nodes within the graph, though the maximum may not be unique. The values of energy modularity range in \( \left[{-1,1}\right[ \).

\subsection{Louvain-based Algorithm}\label{sec:solver}
Maximizing standard modularity is an NP-complete problem~\cite{brandes_modularity_2008}. Energy modularity further complicates the problem as \( d(\mathcal{C}) \) is not static but depends on the community members and their flexibility activation in Eq.~\eqref{eq:d_opt}, which needs to be calculated per community.

\subsubsection{Algorithm}\label{sec:louvain_alg}
This work adapts the well-known Louvain algorithm~\cite{blondel_fast_2008}, a highly scalable, greedy, agglomerative community detection algorithm.
The algorithm maximizes modularity by iteratively applying two phases: \textit{local optimization} and \textit{aggregation}. Each node is initially assigned to its own community, resulting in a singleton partition.
In the local optimization phase, nodes are selected in a random order and evaluated if moving the node to the community of a neighboring node would improve modularity. The best move among the neighboring communities is applied if and only if the move induces a gain in modularity.
This procedure is repeated until no further modularity improvement is possible, whereby the algorithm switches to the aggregation phase to aggregate each community into a virtual super-node. After aggregation, the algorithm starts its second iteration by performing local optimization on a reduced graph based on the virtual super-nodes and their connections, thus reducing the graph size.
This two-phase process repeats until local optimization cannot further increase modularity, yielding the optimal community structure. The resulting partition can vary due to randomness in node reassignment in the local optimization phase.

Pruning is applied in the local optimization phase, reducing the computational time by up to 90\% compared to the original Louvain algorithm (Appendix~\ref{apx:no_pruning}) while retaining quality~\cite{ozaki2016simple}.
Instead of revisiting all nodes to evaluate potential modularity improvements, nodes are evaluated sequentially in a random order only once. When a node is moved, any of its neighbors that belong to a different community and are not in the queue are added and revisited later.

\begin{algorithm}[tb]
    \caption{Louvain algorithm using pruning}\label{alg:louvain}

    \DontPrintSemicolon{}
    \SetKwProg{Fn}{function}{}{end}
    \SetKwBlock{Repeat}{repeat}{}

    \Fn{\( Louvain(\mathcal{G}) \)}{

        \Repeat{
            \( \mathcal{P} = \{\{v\} \;|\; v\in{\mathcal{V}(\mathcal{G})}\} \) \tcp*{Singleton partition}
            \( \mathcal{P}' \gets LocalOptimization(\mathcal{G}, \mathcal{P}) \)\;
            \lIf{\(\mathcal{P} = \mathcal{P}' \)}{
                \textbf{break}
            }
            \( \mathcal{G} \gets Aggregation(\mathcal{G}, \mathcal{P}') \)\;
        }
        \Return{\( \{ flat(\mathcal{C}) \;|\; \mathcal{C} \in{\mathcal{P}} \} \)}
    }
    \BlankLine{}
    \Fn{\( LocalOptimization(\mathcal{G}, \mathcal{P}) \)}{
    \( \mathcal{Q} \gets Queue(\mathcal{V}(\mathcal{G})) \) \tcp*{Randomly ordered}
    \While{\( \mathcal{Q} \neq{} \emptyset{} \)}{
        \( v \gets \mathcal{Q}.pop() \)\;
        \( N \gets \{ \mathcal{C}\in{\mathcal{P}} \;|\; \mathcal{N}_\mathcal{G} (v) \cap \mathcal{C} \neq \emptyset\} \) \;
        \( \mathcal{C}' \gets \argmax_{C \in{}N\cup{}\emptyset}(\Delta Q_{\mathcal{P}}(v \mapsto \mathcal{C})) \) \;
        \uIf{\( \Delta Q_{\mathcal{P}}(v \mapsto \mathcal{C}') > 0 \)}{
            \( v \mapsto \mathcal{C}' \) \;
            \( \mathcal{Q}.add(w) \quad{} \forall{} w \in{}\mathcal{N}_\mathcal{G} (v) \setminus{} (\mathcal{C}' \setminus{} \mathcal{Q}) \) \;
        }
        }
        \Return{\( \mathcal{P} \)}
    }
    \BlankLine{}
    \Fn{\( Aggregation(\mathcal{G}, \mathcal{P}) \)}{
        \( \mathcal{V}' \gets \mathcal{P} \)\;
        \( \mathcal{E}' \gets \{(A,B) \;|\; (u,v) \in{\mathcal{E}(\mathcal{G})}, u \in{A} \in{\mathcal{P}}, v \in{B} \in{\mathcal{P}}\} \)\;
        \Return{\( \mathcal{G} = (\mathcal{V}',\mathcal{E}') \)}
    }
    \BlankLine{}
    \( \mathcal{P} = Louvain(\mathcal{G}) \)
\end{algorithm}

The Louvain algorithm is shown as pseudocode in Algorithm~\ref{alg:louvain} with the notation inspired by~\cite{traag2019louvain}.
The flattening operation used in line 7 to disaggregate the final partition is given for a set \( A \) by \( flat(A) = \bigcup_{a\in{A}} flat(a) \) and \( flat(a) = a \) if \( a \) is not a set itself. The neighborhood of node \( v \) in line 12 is given by \( \mathcal{N}_\mathcal{G} (v) = \{u \;|\; (u,v) \in{\mathcal{E}}\} \).
Further, \( \Delta Q_{\mathcal{P}} (v \mapsto{\mathcal{C}_{j}}) \) in line 13 is written for the gain in modularity when moving node \( v \in \mathcal{C}_{i} \) to a neighboring community \( \mathcal{C}_{j} \) for some partition \( \mathcal{P} \). In particular, only the modularity contribution from the original community \( \mathcal{C}_{i} \) and the target community \( \mathcal{C}_{j} \) are affected because all other communities remain unchanged.
The gain in modularity can be calculated by
\begin{align}\label{eq:delta_q}
    \begin{split}
        \Delta{} Q_{\mathcal{P}} (v \in{}\mathcal{C}_i \mapsto{\mathcal{C}_{j}}) \; = & \; Q^{c} \Bigl(flat (\mathcal{C}_{j}\cup{} \{v\})\Bigr)       \\
        -                                                                             & \; Q^{c} \Bigl(flat (\mathcal{C}_{j})\Bigr)                   \\
        +                                                                             & \; Q^{c} \Bigl(flat (\mathcal{C}_{i} \setminus{} \{v\})\Bigr) \\
        -                                                                             & \; Q^{c} \Bigl(flat (\mathcal{C}_{i})\Bigr) \,
    \end{split}
\end{align}
The communities are flattened so that the node indices within \( Q^c \) correspond to the global node attributes.

\subsubsection{Connected Communities}\label{sec:connectedness}
The Louvain algorithm has a notable limitation: It does not guarantee connectedness of the communities in the final partition~\cite{traag2019louvain}. A potential solution is the Leiden algorithm, which integrates an additional refinement phase between local optimization and aggregation, ensuring the formation of well-connected communities~\cite{traag2019louvain}. Therefore, each locally optimized community is revisited and potentially split if its nodes are not well-connected. For that, the individual edge weights between a node and its community are used. As the edge weights do not represent the community internal connectivity in terms of energy self-sufficiency and flexibility activation influences the self-sufficiency of its community, adapting the Leiden algorithm and its refinement phase is not straightforward.
Therefore, another approach is applied in this work, inspired by the connectedness index as part of the objective function in~\cite{Blumsack2009} and~\cite{Cotilla-Sanchez2013}.
To prevent the formation of disconnected communities, any potential node move that would result in a disconnected community is prohibited by setting \( \Delta{} Q_{\mathcal{P}} (v \in{}\mathcal{C}_i \mapsto{\mathcal{C}_{j}}) = 0 \) if removing \( v \) from \( C_i \) would cause disconnection. Community connectedness can be efficiently verified using tree traversal algorithms, such as depth-first search, by checking whether all nodes in the community are reachable.

\subsubsection{Convergence}\label{sec:convergence}
The combination of the Louvain community detection algorithm to optimize \textit{energy modularity} in Eq.~\eqref{eq:opt_partition} with the community-level self-sufficiency maximization in Eq.~\eqref{eq:d_opt} using flexibility potentials must converge to eventually find a partition. As can be observed from Algorithm~\ref{alg:louvain}, convergence is given if the Louvain iterations over its two phases \textit{LocalOptimization} and \textit{Aggregation} are stopped in line 5 and the function \textit{LocalOptimization} converges in line~16.

Note that, due to \textit{Aggregation} in line 6, the number of nodes of the graph~\( \mathcal{G} \)  strictly decreases with each iteration of the main loop in the \textit{Louvain} algorithm, guaranteeing eventual termination. The termination of the \textit{LocalOptimization} procedure can be understood by observing that each time new nodes are added to the queue \( \mathcal{Q} \) in line~16, the overall modularity \( Q(\mathcal{P}) \) must strictly increase, as ensured by the condition in line~14. Since the number of possible partitions of \( \mathcal{G} \) is finite, the modularity \( Q(\mathcal{P}) \) can only take on finitely many distinct values. As a result, line~16 can be executed only a finite number of times, meaning the queue \( \mathcal{Q} \) will eventually be empty, and the function will terminate. This reasoning holds under the assumption that the computation of \( Q(\mathcal{P}) \) is deterministic, which is indeed the case.

\subsubsection{Scalability}\label{sec:scalability}
The modularity gain is computed quite often in the local optimization phase, therefore, this calculation should be fast. Calculating \( d(\mathcal{C}) \) as part of \( Q^c(\mathcal{C}) \) in Eq.~\eqref{eq:d_opt} can be solved using linear programming.
While solving linear programs is relatively efficient, the performance may still be insufficient for large-scale energy networks or scenarios requiring near-instantaneous results. For instance, in the event of a blackout, the network may need to be partitioned into self-sufficient subnetworks within seconds for a horizon of a few hours up to days.

As an alternative to linear programming, a simulation-based approach can be employed, in which virtual flexibility activation is tracked over time. This method operates under the simplifying assumption that efficiency losses can be neglected, i.e., \( \upeta^\mathrm{p}_{v} = \upeta^\mathrm{u}_{v} = 1 \; \forall v \in \mathcal{V} \) and \( \upeta^\mathrm{f}_{e} = 1 \; \forall e \in \mathcal{E}\). Additionally, energy flow constraints are not taken into account, meaning that Eq.~\eqref{eq:energy_flow_limit} is omitted. Without energy flow efficiency and energy flow limits, the problem can be reduced to an energy balancing problem by aggregating all community nodes into a single node. Similarly, without flexibility usage and preservation efficiency, all flexibility potentials can be aggregated into one representative flexibility.

\begin{algorithm}[tb]
    \caption{Simulation of optimal flexibility activation}\label{alg:flex_contribution_simulation_one_flex}

    \DontPrintSemicolon{}
    \SetNoFillComment{}
    \LinesNotNumbered{}

    \SetKwProg{Fn}{function}{}{end}
    \SetKw{Assumption}{Assumption}
    \SetKw{Input}{Input}

    \Input{\( \mathcal{G}=(\mathcal{V},\mathcal{E}),\,\mathrm{D},\,\mathrm{S},\,\mathrm{\overline{f}},\,\mathrm{\overline{soc}} \)} \tcp*{global parameters}
    \Assumption{\( \upeta^\mathrm{p}_{v} = \upeta^\mathrm{u}_{v} = 1 \; \forall{v\in{\mathcal{V}}}, \, \upeta^\mathrm{f}_{e} = 1 \; \forall{e\in{\mathcal{E}}}\)}\;
    \BlankLine{}

    \Fn{\( SimulateFlex(\mathcal{C}) \)}{
        \( d^{s},\, d^{f} \in{\mathbb{R}} \gets 0 \) \tcp*{demand covered by supply, flex}
        \( \sigma{},\, \sigma^{ub},\, \sigma^{lb} \in{\mathbb{R}} \gets 0 \) \tcp*{virtual SOC and bounds}
        \( \overline{\Delta} \gets \sum_{w \in{\mathcal{C}}} \overline{\mathrm{f}}_{w} \) \tcp*{flex usage limit}
        \( \overline{\sigma} \gets \sum_{w \in{\mathcal{C}}} \overline{\mathrm{soc}}_{w} \) \tcp*{flex capacity limit}
        \For{\( t\in{\mathcal{T}} \)}{
            \( d^{s} \gets d^{s} + \sum_{w \in{\mathcal{C}}} min(\mathrm{S}_{w,t}, \mathrm{D}_{w,t}) \) \;
            \( \Delta \gets \sum_{w \in{\mathcal{C}}} (\mathrm{S}_{w,t} - \mathrm{D}_{w,t}) \) \;
            \( \Delta \gets \max\Bigl(\min\bigl(\Delta\,,\, \mathrm{\overline{\Delta}}\bigr)\,,\, -\mathrm{\overline{\Delta}}\Bigr) \) \tcp*{flex usage limit}
            \uIf(\tcp*[f]{oversupply: store energy}){\( \Delta > 0 \)}{
                \( \sigma \gets \min\bigl(\sigma + \Delta  \,,\, \sigma^{lb} + \mathrm{\overline{\sigma}}\bigr) \) \tcp*{flex capacity limit}
                \( \sigma^{ub} \gets \max\bigl( \sigma^{ub},\, \sigma\bigr) \) \tcp*{update upper bound}
            }
            \Else(\tcp*[f]{undersupply: dispatch energy}){
                \( \sigma^{old} \gets \sigma \) \;
                \( \sigma \gets \max\bigl(\sigma - \Delta  \,,\,\sigma^{ub} - \mathrm{\overline{\sigma}}\bigr) \) \tcp*{flex capacity limit}
                \( \sigma^{lb} \gets \min\bigl(\sigma^{lb},\, \sigma\bigr) \) \tcp*{update lower bound}
                \( d^{f} \gets d^{f} + (\sigma^{old} - \sigma) \) \;
            }
        }
        \( d^{f} \gets{} d^{f}+ \min(\sigma, 0) \)\tcp*{compensate over-dispatch}
        \Return{\( d^{s} + d^{f}\)}
    }

    \BlankLine{}
    \( d(\mathcal{C}) = SimulateFlex(\mathcal{C})\) \tcp*{\( \mathcal{C} \) is flat}
\end{algorithm}

The simulation-based algorithm is described as pseudocode in Algorithm~\ref{alg:flex_contribution_simulation_one_flex}. Within a community \( \mathcal{C} \), presumed flexibility activation is tracked over time \( t \) to derive the total self-sufficiency \( d(\mathcal{C}) \).
At the start, the virtual state of charge of the aggregated flexibility resource is zero. Positive and negative flexibility can be provided until upper and lower bounds are reached.
In case of oversupply in the community, energy is stored, limited by Eq.~\eqref{eq:f_max}, and the virtual state of charge is increased, following the constraints on the state of charge changes over time in Eq.~\eqref{eq:SOC_time}. Undersupply is handled the opposite way in addition to tracking flexibility activation. Since the initial state of charge is assumed, upper and lower state of charge bounds are tracked to comply with Eq.~\eqref{eq:SOC_minmax}. Those bounds are updated in each \( t \) and respected when updating the virtual state of charge. Subsequently, the next time slice \( t + 1 \) is processed. To comply with the cyclic state of charge constraint in Eq.~\eqref{eq:SOC_cyclic}, over-dispatch is compensated by subtracting the final virtual state of charge from the tracked flexibility activation.

The computational complexity of calculating \( Q^c(\mathcal{C}) \) using the simulation-based approach is \( \mathcal{O}\bigl(|\mathcal{C}| \cdot |\mathcal{T}|\bigr) \), where \( |\mathcal{C}| \) denotes the size of the community and \( |\mathcal{T}| \) the length of the time horizon.
In the first iteration of the Louvain algorithm, each node is initially assigned to its own community, resulting in \( |\mathcal{C}| = 1 \). This number increases in subsequent iterations as nodes are merged. The theoretical worst-case scenario, in which all nodes are grouped into a single community, only occurs with a resolution parameter \( \gamma = 0 \), a setting that is unsuitable for meaningful community detection.

The computational complexity of verifying whether the resulting communities from \( \Delta Q_{\mathcal{P}}(v \in \mathcal{C}_i \mapsto \mathcal{C}_j) \) remain connected is \( \mathcal{O}\bigl(|\mathcal{C}_i| + |\mathcal{E}'_i|\bigr) \), where \( \mathcal{E}'_i \) denotes the set of internal edges within community \( \mathcal{C}_i \). This is significantly lower in scale than the complexity of calculating \( Q^c(\mathcal{C}) \), and can therefore be considered negligible in the scalability analysis.
\section{Numerical Evaluation}\label{sec:evaluation}
This section evaluates the proposed algorithm using an electrical power grid as a typical energy network.
The calculation of \( d(\mathcal{C}) \) in Eq.~\eqref{eq:d_opt} is carried out using three distinct approaches:
(1) \textit{NoFlex} assumes lossless energy flows by considering only the basic energy balance, while completely disregarding flexibility potentials.
(2) \textit{SimulateFlex} is the simulation-based method, detailed in Algorithm~\ref{alg:flex_contribution_simulation_one_flex}, which incorporates flexibility potentials but omits efficiencies and flow constraints.
(3) \textit{LPFlex} employs a linear programming formulation that captures flexibility potentials, accounts for all relevant efficiencies, and enforces energy flow limit constraints.

\subsection{Test System Data}
\begin{figure}[tb]
    \centering{}
    \includegraphics[width=\linewidth]{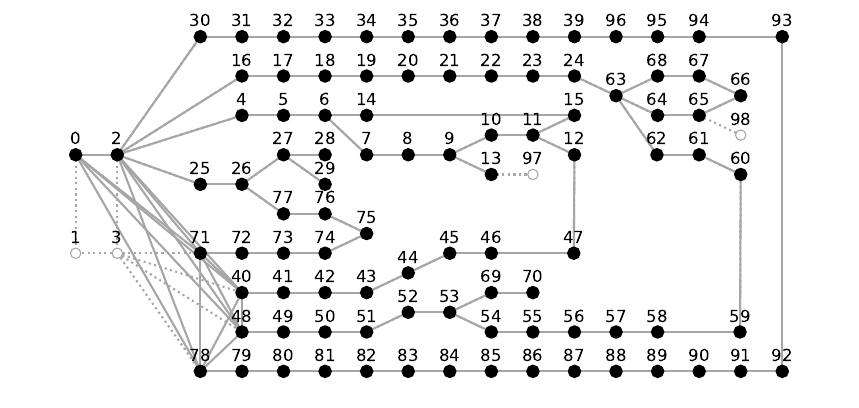}
    \caption{SimBench Grid \texttt{``1-MV-rural--1-sw''}, white nodes have neither demand nor supply and are therefore removed}
    \label{fig:simbench_grid}
    \Description{Plot of the selected SimBench power grid, showing nodes, node identifiers and edges.}
\end{figure}
For reproducibility, the publicly available Simbench dataset~\cite{meinecke2020simbench} is used. Shown in Figure~\ref{fig:simbench_grid}, a representative 99-bus medium voltage rural power grid with 105 bidirectional edges (lines and transformers) is selected, and asset availability and behavior are projected for 2024 (Simbench code \texttt{1-MV-rural-{}-1-sw}). Generators are attached to 94 buses, energy demands to 92 buses. The grid includes simulated demand and supply profiles in quarter-hourly resolution for an entire year. One node represents the upstream grid; the remaining are low-voltage grids with aggregated supply and demand profiles, industrial-scale consumers, wind, photovoltaic, biomass, or hydropower generators. Nodes 0 and 1 connect to the high-voltage upstream grid; Nodes 2 and 3 are busbars with directly connected resources at node 2.

Overall, the grid is a net-generating grid with an annual demand of 32.25~GWh and an annual supply of 58.44~GWh. Energy storage systems are installed at 53 buses with 12.57~MWh of capacity that can provide positive and negative flexibility up to 1.57~MWh per quarter-hour. Flexibility resource efficiencies are \( \upeta^\mathrm{u}_{v} = 0.95\), \( \upeta^\mathrm{p}_{v} = 0.9986,\, \forall{v\in{\mathcal{V}}} \), and edge efficiency is \( \upeta^\mathrm{f}_{e} = 0.95,\, \forall{e\in{\mathcal{E}}} \).
Switches to connect open rings are considered closed. Nodes without demand, supply, or flexibility, like busbars or coupling points, are removed (node 1, 3, 97, 98). Their neighbors are connected with new edges to preserve information on the capability of exchanging energy between them. The energy network without flexibility resources has an energy self-sufficiency of 82.1\% with \textit{LPFlex} (84.3\% with \textit{NoFlex}) and an energy self-sufficiency when considering flexibility resources of 88.3\% with \textit{LPFlex} (90.9\% with \textit{SimulateFlex}). Energy supply from the upstream grid is simulated by a supply profile attached as generator to the slack bus 0, covering remaining demand without using flexibility resources.

Each experiment is conducted 30 times to assess the impact of randomness in the Louvain algorithm. Because of the randomness involved, it is reasonable to execute the algorithm several times and select the partition which fits the network best.
The algorithms are implemented in Python and linear programming is solved in Gurobi version 12.0.1. Experiments are run on Intel Xeon E5-2650v2 @ 2.60 GHz with 32~GB RAM;\@ the limiting factor is the processing power. Results on runtime performance should be considered as a point of reference with further improvement possibilities when using a performance-oriented programming language.

\subsection{Scalability: NoFlex, SimulateFlex, LPFlex}\label{sec:eval_lp_sim}
To assess the scalability of the proposed approach, first the runtime of \textit{NoFlex}, \textit{SimulateFlex} and \textit{LPFlex} to calculate \( d(\mathcal{C}) \) is evaluated. Then, the runtime of the overall Louvain algorithm with increasing horizon is analyzed.

\subsubsection{Calculating Self-Sufficiency}
To empirically evaluate the scalability of calculating \( d(\mathcal{C}) \), both \( \mathcal{C} \) (spatial) and \( \mathcal{T} \) (temporal) are scaled independently to measure their impact on the runtime for \textit{NoFlex}, \textit{SimulateFlex} and \textit{LPFlex}. This experiment is only run once, as no randomness is involved.

\begin{figure}[tb]
    \centering{}
    \includegraphics[width=\linewidth]{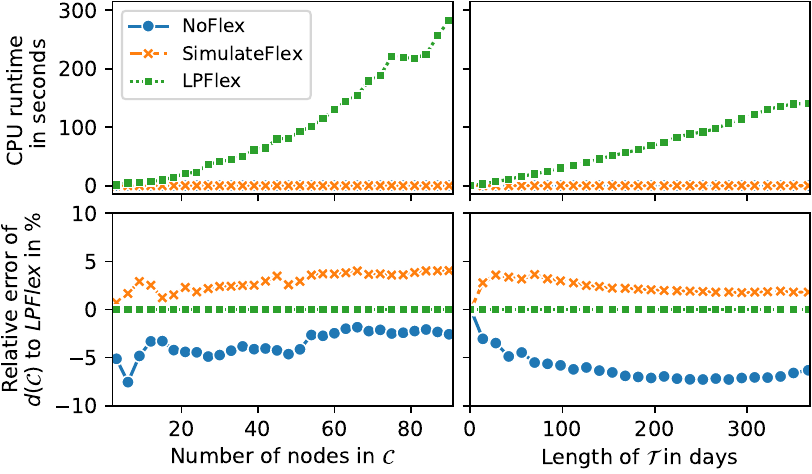}
    \caption{Runtime and accuracy of different methods to approximate \( d(\mathcal{C}) \) with varying \( \mathcal{C} \) and \( \mathcal{T} \)}
    \label{fig:scaling_flex_opt}
    \Description{This figure shows the scaling of the three different approaches \textit{NoFlex}, \textit{SimulateFlex} and \textit{LPFlex} to calculate \( d(\mathcal{C}) \).}
\end{figure}

On the left side of the Figure~\ref{fig:scaling_flex_opt}, the number of time steps is fixed to \( |\mathcal{T}| = 30 \) days, while the number of nodes in \( \mathcal{C} \) is varied from 3 to 90. The runtime of the \textit{LPFlex} approach increases non-linearly with the number of nodes. This is primarily due to the growing number of variables in the underlying linear program, which scales not only with the number of nodes in \( \mathcal{C} \) but also with the associated network edges \( \mathcal{E} \) and their corresponding energy flow variables \( W_{e,t} \). In comparison, both \textit{SimulateFlex} and \textit{NoFlex} are by an order of magnitude faster and scale linear with \( \mathcal{C} \).

The relative error in the resulting self-sufficiency values \( d(\mathcal{C}) \) is shown with respect to the \textit{LPFlex} as baseline. \textit{SimulateFlex} overestimates self-sufficiency, as it neglects efficiency losses in both flexibility activation and energy flows. In contrast, \textit{NoFlex} underestimates self-sufficiency because it entirely disregards flexibility potentials. The relative error of \textit{SimulateFlex} slightly increases with the number of nodes in \( \mathcal{C} \), as this results in more energy flows and consequently greater deviations due to omitted flow efficiency.

On the right side of the Figure~\ref{fig:scaling_flex_opt}, the spatial resolution is fixed at \( |\mathcal{C}| = 10 \) nodes, while the time set \( \mathcal{T} \) is increased from a single day to a full year, i.e., 35136 quarter-hour time slices. In this case, runtime scales almost linearly with the time set length for all approaches. However, the \textit{LPFlex} approach shows a much steeper increase in runtime compared to \textit{SimulateFlex}, again reflecting the higher complexity of solving the linear program.
For a full year of data using the complete energy network with \( \mathcal{C} = \mathcal{V} \), the \textit{LPFlex} approach required approximately 2.5 hours of computation time while \textit{SimulateFlex} only takes 0.04 seconds.

Similar to the spatial scaling scenario, \textit{SimulateFlex} consistently overestimates, while \textit{NoFlex} underestimates the self-sufficiency \( d(\mathcal{C}) \). Despite this, the relative error of \textit{SimulateFlex} remains within an acceptable range, indicating that the simplified simulation approach captures the essential system dynamics in temporal scale while offering significantly lower computational effort.

\subsubsection{Louvain with Varying Horizon}
\begin{figure}[tb]
    \centering{}
    \includegraphics[width=\linewidth]{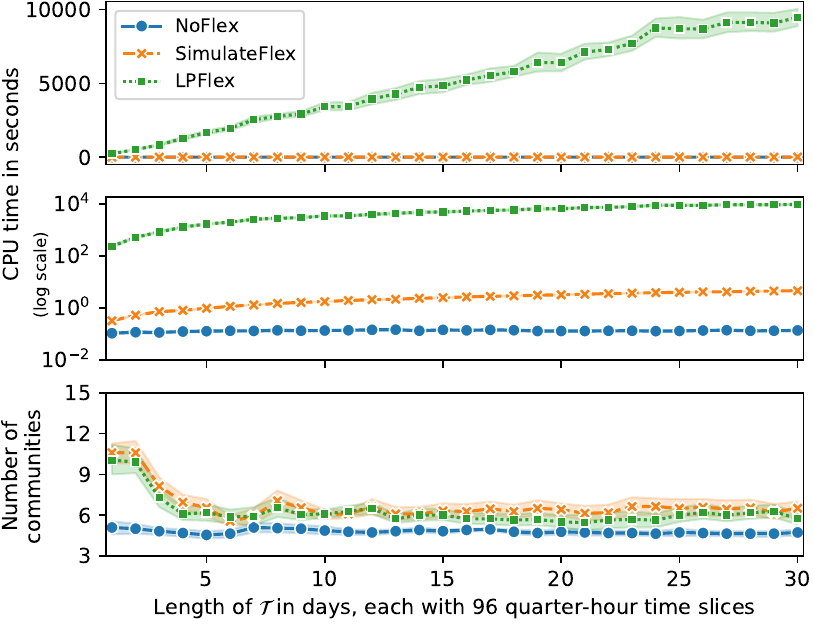}
    \caption{Louvain runtime with increasing \( \mathcal{T} \) and \( \gamma = 0.3\), 0.95 confidence intervals}
    \label{fig:scaling_t_runtime}
    \Description{This figure shows the scaling of the Louvain algorithm with increasing time horizon.}
\end{figure}

To evaluate the runtime behavior of the complete Louvain algorithm with respect to the temporal resolution, various length of the time set \( \mathcal{T} \) were analyzed, ranging from a single day to thirty consecutive days, starting on April 1\textsuperscript{st}. The resolution parameter was fixed at \( \gamma = 0.3 \). The corresponding results are presented in Figure~\ref{fig:scaling_t_runtime}.

The runtime of the Louvain algorithm using the \textit{NoFlex} approach remains constant across different lengths of the time set \( \mathcal{T} \). In contrast, the runtime of the \textit{LPFlex} approach increases almost linearly with the size of \( \mathcal{T} \), but with a substantially steeper slope compared to the \textit{SimulateFlex} approach.
Although \textit{LPFlex} demonstrates non-linear scalability with respect to the number of nodes in a community, this effect is effectively mitigated within the iterative structure of the Louvain algorithm. Because Louvain starts with each node in its own community, \textit{LPFlex} is applied frequently to small communities and less often to increasingly aggregated ones. As a result, the impact of non-linear scaling with increasing \( \mathcal{C} \) is significantly reduced, and runtime scalability remains almost linear. However, running the Louvain with \textit{LPFlex} does not scale well.

In contrast, the linear runtime trend of simulation-based approach \textit{SimulateFlex} persists even when scaling up to a full year, consisting of 35,136 quarter-hours. In this scenario, \textit{SimulateFlex} completed in 41.4 seconds, underscoring its computational efficiency and suitability for large-scale analysis.

The modularity remains relatively stable with increasing size of the time set \( \mathcal{T} \). However, as shown in the lower part of Figure~\ref{fig:scaling_t_runtime}, a higher number of communities is identified for shorter time horizons, less than five days, when flexibility is considered.

\subsection{Impact of the Resolution Parameter}\label{sec:eval_gamma}
\begin{figure}[tb]
    \begin{subfigure}{\linewidth}
        \includegraphics[width=\linewidth]{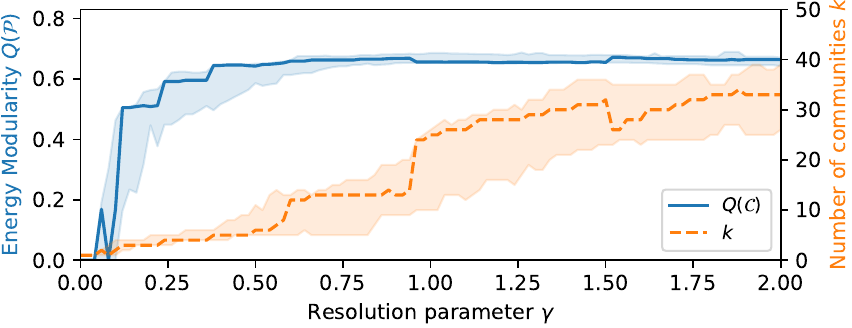}
        \caption{NoFlex: Energy balance, ignoring flexibility}
        \label{fig:resolution_no_flex}
    \end{subfigure}
    \begin{subfigure}{\linewidth}
        \includegraphics[width=\linewidth]{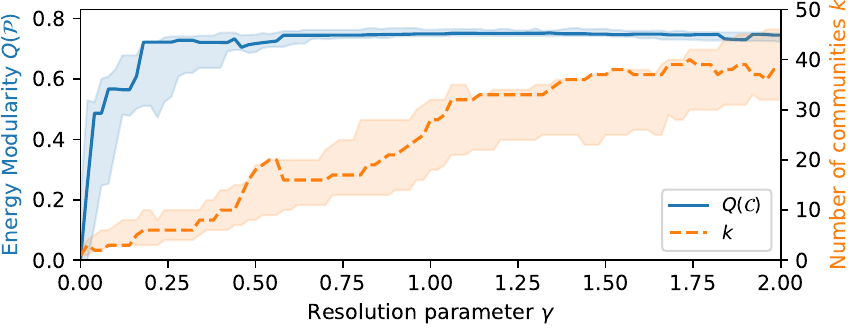}
        \caption{SimulateFlex: Energy balance, ignoring flexibility efficiencies}
        \label{fig:resolution_flex}
    \end{subfigure}
    \caption{Varying \( \gamma \) (in 0.02 steps), areas show min and max of 30 runs, the lines show a selected run}
    \label{fig:resolution}
    \Description{This figure shows the effect of varying resolution parameter \( \gamma{} \).}
\end{figure}

One month of data is used to analyze the impact of the resolution parameter \( \gamma \). April has been selected with average outside temperatures and mixed weather. \( \gamma \) values are tested in steps of 0.02 between 0 and 2. The energy modularity of the final partition is recalculated with \( \gamma = 1 \) and plotted together with the number of communities \( k \). \textit{NoFlex}, i.e., ignoring flexibility, is applied in Figure~\ref{fig:resolution_no_flex}, and \textit{SimulateFlex} is applied in Figure~\ref{fig:resolution_flex}.

Figure~\ref{fig:resolution} shows that energy modularity is lower with \textit{NoFlex} compared to \textit{SimulateFlex}. This observation is expected, as only existing flexibility could improve self-sufficiency as a central part of energy modularity.
Furthermore, the number of detected communities is slightly higher when using flexibility. This observation follows intuition: Undersupply periods, like during the night, can be reduced with flexibility activation, which shifts energy from oversupply periods, like during a sunny day. This flexibility activation improves the self-sufficiency of the given community until it reaches 100\%. A higher number of self-sufficient communities tends to increase the energy modularity, since the subtractive term in Eq.~\eqref{eq:q_c} becomes smaller, leading to more communities when using flexibility.

Furthermore, it can be observed that energy modularity increases with increasing \( \gamma \) up to a certain point around \( \gamma \approx 0.75 \). With \( \gamma > 1 \), energy modularity seems almost constant but is slightly decreasing. In the given network, all nodes have energy supply, and almost all also have demand. Thus, even with a singleton partition with each node in its own community, the average self-sufficiency is significantly higher than zero but lower than when forming communities.
Lower \( \gamma \) values converge to fewer communities and larger \( \gamma \) to more communities. There is a sweet spot with \( \gamma \) values around 0.2 - 0.5, where the energy modularity is high and the number of resulting communities is small. From the definition of energy modularity in Eq.~\eqref{eq:q_c}, it is known that smaller \( \gamma \) emphasizes the connectivity within a community. Thus, smaller \( \gamma \) emphasizes energy self-sufficiency, which can be seen as preferred if the overall energy modularity is not significantly smaller.

Applying the community detection algorithm to a full year of data results in a similar impact of \( \gamma \). However, the overall energy modularity is slightly lower because long-term flexibility resources are unavailable for shifting oversupply from photovoltaic generation from summer to winter, thus decreasing the annual self-sufficiency.

\begin{figure}[tb]
    \centering{}
    \includegraphics[width=\linewidth]{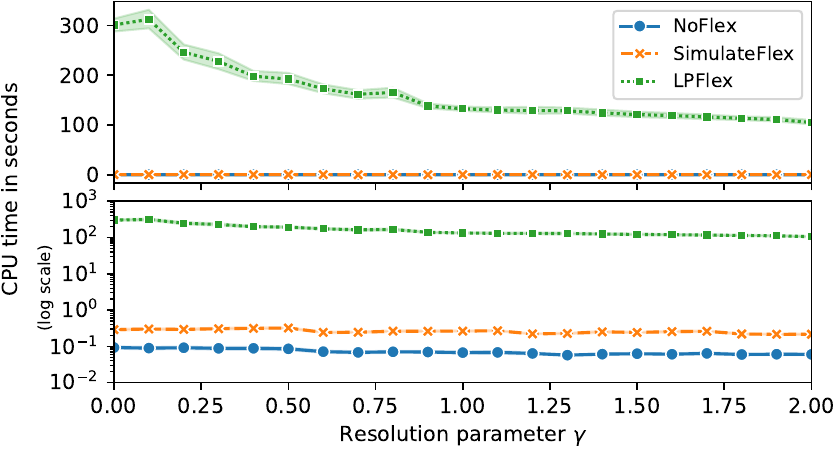}
    \caption{Runtime of Louvain algorithm with varying \( \gamma \) (in 0.1 steps) and 0.95 confidence intervals for one day (\( |\mathcal{T}| = 96 \))}
    \label{fig:resolution_runtime}
    \Description{This figure shows the runtime of the Louvain algorithm using the three different approaches \textit{NoFlex}, \textit{SimulateFlex} and \textit{LPFlex}.}
\end{figure}

Processing times with varying \( \gamma \) in 0.1 steps and using a single day of data, i.e., 96 intervals in \( \mathcal{T} \), are shown in Figure~\ref{fig:resolution_runtime} for \textit{NoFlex}, \textit{SimulateFlex} and \textit{LPFlex}. It can be observed that \textit{SimulateFlex} achieves significantly faster runtimes, with an average of 0.26 seconds compared to \textit{LPFlex}, with an average of 132 seconds. Furthermore, all methods exhibit slightly longer processing times for smaller \( \gamma \) values, linked to fewer communities with more members. This effect is particularly pronounced in the case of \textit{LPFlex} where the runtime at \( \gamma = 0.1 \) is approximately twice as high as at \( \gamma = 1 \).

\subsection{Partition Quality}
The solution of the Louvain algorithm depends on the initially randomly ordered queue of nodes. While every execution with a different random seed yields a similar partition in terms of quality, like its modularity or number of communities, the actual partition might differ. This is especially the case when node characteristics are similar and many similarly good partitions exist. In the given example network, all runs produced good partitions with different layouts. Thus, the following results are considered one of many partitions with greedily maximized energy modularity. For comparison, the same random seed has been used.
One month of data is used (April) with \( \gamma = 0.25 \) in the sweet spot of Figure~\ref{fig:resolution}.

\begin{figure}[tb]
    \centering{}
    \includegraphics[width=\linewidth]{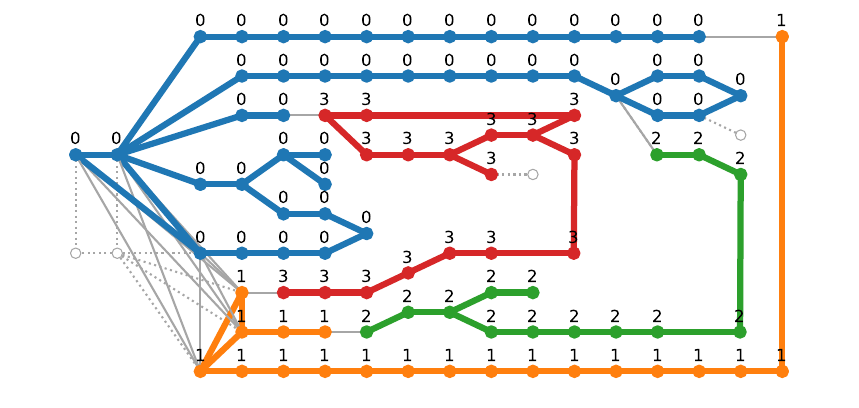}
    \caption{Partition with \textit{NoFlex}}
    \label{fig:example_noflex}
    \Description{Plot of the partition of the energy network using \textit{NoFlex}}
\end{figure}
\begin{table}[tb]
    \caption{Partition with \textit{NoFlex}}
    \label{tab:stats_noflex}
    \small{}
    \centering{}
    \begin{tabular}{c c c c c}
        \toprule{}
        ID      & \shortstack[c]{Number of                                                                                                                                                                                                                             \\ members}  & Demand & \shortstack[c]{Self-sufficiency \\ (\textit{NoFlex})} & \shortstack[c]{Self-sufficiency \\ (\textit{LPFlex})} \\
        \( i \) & \( |\mathcal{C}| \)      & \( a (\mathcal{C}) \) & \( \frac{d (\mathcal{C})}{\sum_{t\in{\mathcal{T}}}\sum_{w\in{\mathcal{C}}} \mathrm{D}_{w,t}} \) & \( \frac{d (\mathcal{C})}{\sum_{t\in{\mathcal{T}}}\sum_{w\in{\mathcal{C}}} \mathrm{D}_{w,t}} \) \\
        \midrule{}
        0       & 44                       & 49.3\%                & 92.5\%                                                                                          & 98.1\%                                                                                          \\
        1       & 20                       & 17.0\%                & 86.1\%                                                                                          & 95.0\%                                                                                          \\
        2       & 14                       & 19.3\%                & 81.8\%                                                                                          & 84.2\%                                                                                          \\
        3       & 17                       & 14.5\%                & 93.3\%                                                                                          & 99.4\%                                                                                          \\
        \bottomrule{}
    \end{tabular}
\end{table}

\begin{figure}[tb]
    \centering{}
    \includegraphics[width=\linewidth]{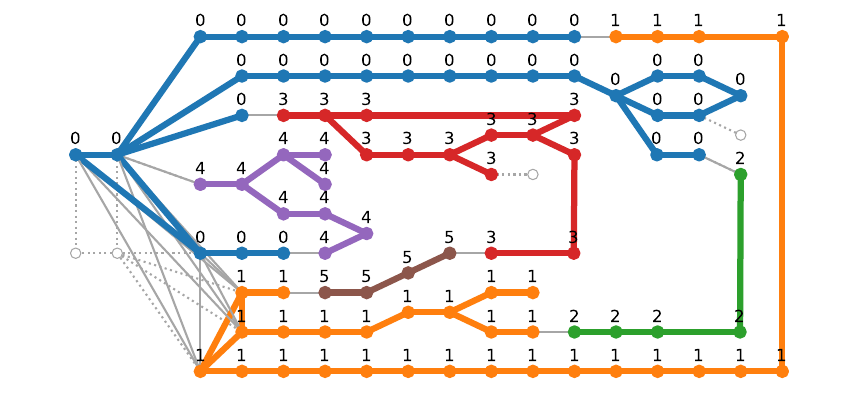}
    \caption{Partition with \textit{LPFlex}}
    \label{fig:example_simulateflex}
    \Description{Plot of the partition of the energy network using \textit{LPFlex}}
\end{figure}
\begin{table}[tb]
    \caption{Partition with \textit{LPFlex}}
    \label{tab:stats_simulateflex}
    \small{}
    \centering{}
    \begin{tabular}{c c c c c}
        \toprule{}
        ID      & \shortstack[c]{Number of                                                                                                                                                                                                                             \\ members}  & Demand & \shortstack[c]{Self-sufficiency \\ (\textit{NoFlex})} & \shortstack[c]{Self-sufficiency \\ (\textit{LPFlex})} \\
        \( i \) & \( |\mathcal{C}| \)      & \( a (\mathcal{C}) \) & \( \frac{d (\mathcal{C})}{\sum_{t\in{\mathcal{T}}}\sum_{w\in{\mathcal{C}}} \mathrm{D}_{w,t}} \) & \( \frac{d (\mathcal{C})}{\sum_{t\in{\mathcal{T}}}\sum_{w\in{\mathcal{C}}} \mathrm{D}_{w,t}} \) \\
        \midrule{}
        0       & 33                       & 36.1\%                & 91.0\%                                                                                          & 96.4\%                                                                                          \\
        1       & 31                       & 35.1\%                & 82.5\%                                                                                          & 89.5\%                                                                                          \\
        2       & 5                        & 4.7\%                 & 88.0\%                                                                                          & 97.9\%                                                                                          \\
        3       & 13                       & 11.5\%                & 93.3\%                                                                                          & 98.6\%                                                                                          \\
        4       & 9                        & 9.5\%                 & 86.9\%                                                                                          & 97.8\%                                                                                          \\
        5       & 4                        & 3.0\%                 & 74.3\%                                                                                          & 99.8\%                                                                                          \\
        \bottomrule{}
    \end{tabular}
\end{table}

\begin{figure}[tb]
    \centering{}
    \includegraphics[width=\linewidth]{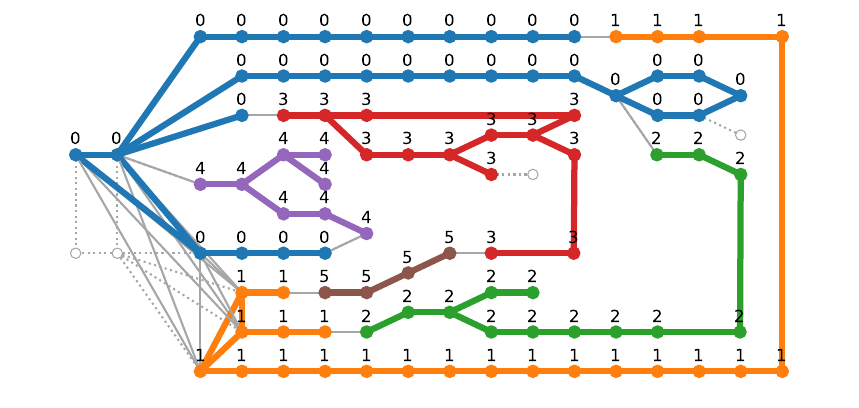}
    \caption{Partition with \textit{SimulateFlex}}
    \label{fig:example_lpflex}
    \Description{Plot of the partition of the energy network using \textit{SimulateFlex}}
\end{figure}
\begin{table}[tb]
    \caption{Partition with \textit{SimulateFlex}}
    \label{tab:stats_lpflex}
    \small{}
    \centering{}
    \begin{tabular}{c c c c c}
        \toprule{}
        ID      & \shortstack[c]{Number of                                                                                                                                                                                                                             \\ members}  & Demand & \shortstack[c]{Self-sufficiency \\ (\textit{NoFlex})} & \shortstack[c]{Self-sufficiency \\ (\textit{LPFlex})} \\
        \( i \) & \( |\mathcal{C}| \)      & \( a (\mathcal{C}) \) & \( \frac{d (\mathcal{C})}{\sum_{t\in{\mathcal{T}}}\sum_{w\in{\mathcal{C}}} \mathrm{D}_{w,t}} \) & \( \frac{d (\mathcal{C})}{\sum_{t\in{\mathcal{T}}}\sum_{w\in{\mathcal{C}}} \mathrm{D}_{w,t}} \) \\
        \midrule{}
        0       & 32                       & 35.3\%                & 88.6\%                                                                                          & 94.8\%                                                                                          \\
        1       & 24                       & 22.6\%                & 83.4\%                                                                                          & 93.2\%                                                                                          \\
        2       & 14                       & 19.3\%                & 81.8\%                                                                                          & 84.2\%                                                                                          \\
        3       & 13                       & 11.5\%                & 93.3\%                                                                                          & 98.6\%                                                                                          \\
        4       & 8                        & 8.3\%                 & 87.8\%                                                                                          & 98.4\%                                                                                          \\
        5       & 4                        & 3.0\%                 & 74.3\%                                                                                          & 99.8\%                                                                                          \\
        \bottomrule{}
    \end{tabular}
\end{table}

\subsubsection{Partition with \textit{NoFlex} vs.\ \textit{SimulateFlex}}
Four communities of different sizes are detected when using \textit{NoFlex}, as shown in Figure~\ref{fig:example_noflex}. Statistics of the communities are listed in Table~\ref{tab:stats_noflex}. The overall self-sufficiency when ignoring flexibility ranges above 81\%, while the self-sufficiency with flexibility is slightly higher.
When considering flexibility resources, six communities are detected; see Figure~\ref{fig:example_simulateflex}. As can be observed from Table~\ref{tab:stats_simulateflex}, the self-sufficiency is relatively high for all detected communities. It can be noted that when integrating flexibility potentials into the community detection, the number of resulting communities increased with a similar self-sufficiency rate. Thus, it yields more advanced communities that use flexibility to increase their self-sufficiency.
Furthermore, community 5, which achieves only 74\% self-sufficiency without flexibility, reaches 99.8\% self-sufficiency when flexibility is utilized. This highlights the substantial impact of flexibility on the resulting partition and demonstrates its critical role in enhancing energy autarky.

\subsubsection{Partition with \textit{LPFlex} vs.\ \textit{SimulateFlex}}
When comparing the Partitions when using \textit{LPFlex} versus \textit{SimulateFlex} with the same random seed, it can be observed that the core of most communities are the same in Figure~\ref{fig:example_lpflex}. Also, the self-sufficiency of the resulting communities is quite similar, cf. Table~\ref{tab:stats_lpflex}. As expected, neglecting the flexibility usage efficiency has only a minor impact on the resulting partition.

\subsection{Discussions and Limitations}\label{sec:discussion}
Energy modularity is an effective metric for evaluating community structures, focusing on self-sufficiency. By integrating simulation-based flexibility potential calculation, i.e., \textit{SimulateFlex}, into the Louvain algorithm for community detection, fast execution times with reasonable accuracy are achieved. This efficiency enables temporal high-resolution analysis of energy network structures while incorporating optimal flexibility activation on community-level.

As \textit{SimulateFlex} assumes a lossless energy network, it tends to overestimate self-sufficiency, meaning that complete self-sufficiency within the resulting communities may not be fully achievable in practice. Nevertheless, actual network losses are expected to be small, especially when energy flows remain predominantly local or regional.

Regarding practical implementation challenges, the proposed methodology requires time series data for demand and supply and metadata to model the constraints on flexibility potentials. This data can be obtained using standard load profiles, as done in this analysis, or through measurements from advanced metering infrastructure.
To enable the actual disconnection of the identified communities, technical measures for operating the community energy network, such as grid-forming resources in power systems, must be in place. Additionally, technical constraints that are not considered in the community detection process, such as voltage stability, reactive power flows, quadratic line losses, and capacity limits, should be evaluated in a subsequent analysis focused on the operational feasibility of the community energy subnetwork.

\section{Conclusions}\label{sec:conclusion}
This work presents an approach to community detection in energy networks that focuses on the self-sufficiency of resulting communities covering both temporal and spatial domains. The temporal domain is defined by energy demand and supply profiles. The spatial domain is modeled as a graph in which edges represent the capability to exchange energy. Combining both with flexibility resources yields the challenge that the actual energy flows on the edges and the self-sufficiency of communities depend on the flexibility activation, which further depends on its assignment to a community. In order to break this cyclic dependency the presented approach focuses on energy-self-sufficiency instead of energy flows.

To this end, a novel \textit{energy modularity} metric, conceptually inspired by standard modularity, is introduced and embedded within the greedy, agglomerative Louvain community detection algorithm. To evaluate community self-sufficiency as part of energy modularity under flexibility potentials, three alternative methods are proposed and compared: (i) ignoring flexibility, (ii) optimizing flexibility activation via linear programming, and (iii) simulating flexibility activation while omitting efficiency losses.

The proposed methodology effectively partitions energy networks into self-sufficient communities, leveraging available flexibility. The evaluation on a representative benchmark power grid demonstrated both the practical applicability and computational efficiency of the approach. The results highlight the potential of energy modularity as a meaningful and operationally relevant metric for community detection focusing on energy self-sufficiency and capturing the complex temporal and spatial interplay between renewable generation, demand, and flexibility.

Future work will aim to refine the physical modeling of various types of energy networks to better reflect real-world operational constraints. Moreover, as an alternative to the greedy Louvain algorithm, a mixed-integer linear programming (MILP) formulation, encoding community membership as a binary decision matrix, will be explored for its potential to yield globally optimal partitions~\cite{miyauchi_exact_2018}.

\appendix{}

\section{Louvain Algorithm: Optimization Phase}\label{apx:no_pruning}
The original Louvain algorithm, described in~\cite{blondel_fast_2008}, uses the following algorithm in the \textit{LocalOptimization} function (Pseudocode in Algorithm~\ref{alg:louvain_local_opt_original}): As long as the overall modularity improves, each node is revisited in a random order to evaluate whether moving it to a neighboring community would further improve modularity. If a node move only occurs in a part of the graph, nodes that are already part of their best surrounding community would be unnecessarily re-evaluated. Pruning is proposed in~\cite{ozaki2016simple} to avoid such situations and speed up the local optimization phase.

\begin{algorithm}[!htb]
    \caption{Local Optimization without pruning}\label{alg:louvain_local_opt_original}

    \DontPrintSemicolon{}
    \SetNoFillComment{}
    \LinesNotNumbered{}
    \SetKwProg{Fn}{function}{}{end}
    \SetKwRepeat{Do}{do}{while}

    \Fn{\( LocalOptimization(\mathcal{G}, \mathcal{P}) \)}{
    \Do{\( Q(\mathcal{P}) > Q_{old} \)}{
        \( Q_{old} \gets Q(\mathcal{P}) \)\;
        \ForEach(\tcp*[f]{Random order}){\( v \in \mathcal{V}(\mathcal{G}) \)}{
            \( N \gets \{ \mathcal{C}\in{\mathcal{P}} \;|\; \mathcal{N}_\mathcal{G} (v) \cap \mathcal{C} \neq \emptyset\} \) \;
            \( \mathcal{C}' \gets \argmax_{C \in{}N\cup{}\emptyset}(\Delta Q_{\mathcal{P}}(v \mapsto \mathcal{C})) \) \;
            \uIf{\( \Delta Q_{\mathcal{P}}(v \mapsto \mathcal{C}') > 0 \)}{
                \( v \mapsto \mathcal{C}' \) \;
            }
            }
        }
        \Return{\( \mathcal{P} \)}
    }
\end{algorithm}

\begin{acks}
  This project has received funding from the European Union's Horizon 2020 research and innovation programme under grant agreement No 957819. Further, the authors want to thank Julian Danner for feedback and discussions.
\end{acks}

\bibliographystyle{ACM-Reference-Format}
\bibliography{bib}

\end{document}